\documentclass[10pt,aps,twocolumn,superscriptaddress,preprintnumbers,nofootinbib]{revtex4-2}
\usepackage{epsfig}  
\usepackage{hyperref}
\usepackage{color}
\usepackage{float}
\usepackage{amsfonts}
\usepackage{amsmath}
\usepackage{slashed}
\usepackage{float}
\usepackage{graphicx}
\usepackage{verbatim}
\usepackage{xcolor}
\usepackage{caption}
\usepackage{physics}
\usepackage{subcaption}
\usepackage{lipsum}
\usepackage{ulem}
\usepackage{import}
\usepackage{array}

\large

\preprint{}
\begin{document}
\title{Model-independent analysis of new physics effects in $B\rightarrow K^*_2(1430)\mu^+ \mu^-$ decay}

\author{Ria Sain}
\email{ria.sain.2013@gmail.com}
\affiliation{Department of Physics, Indian Institute of Technology Guwahati, Assam 781039, India}

\author{Juhi Vardani}
\email{vardani.1@iitj.ac.in}
\affiliation{Indian Institute of Technology Jodhpur, Jodhpur 342037, India}

\begin{abstract}
Recently, the LHCb Collaboration provided updated measurements for the lepton flavour ratios $R_K$ and $R_{K^*}$. The currently observed values align with the predictions of the standard model.  In light of these recent updates, our investigation delves into the repercussions of new physics characterized by universal couplings to electrons and muons. We specifically focus on their impact on various observables within the $B\to K_2^*(1430)(\to K\pi)\mu^+ \mu^-$ decay. These observables include the differential branching ratio, forward-backward asymmetry ($A_{FB}$), longitudinal polarization asymmetry ($F_L$), and a set of optimized observables ($P_i$). Our findings indicate that the branching ratio of $B\to K_2^*(\to K\pi)\mu^+ \mu^-$ decay can be suppressed up to $25\%$ for various new physics solutions. Furthermore, all permissible new physics scenarios demonstrate finite enhancement in the muon forward-backward asymmetry $(A_{FB})$ as well as an increase in the value of the optimized angular observable $P_2$. Moreover, in the presence of new physics zero crossing points for $A_{FB}$ and $P_2$ shift towards higher $q^2$. The current data have a mild deviation from SM predictions in  $P_5'$ observable in the low-$q^2$ bin.  We also explored massive $Z'$ models, which can generate universal 1D new physics scenarios, characterized by  $C_9^{NP}<0$, $C_9^{NP}=-C_{10}^{NP}$, and $C_9^{NP}=-C_9'$. Using additional constraints coming from $B_s-\overline{B_s}$ mixing and neutrino trident process, we find that the conclusions of the model-independent analysis remain valid.

\end{abstract}
 
\maketitle 
\newpage
\section{Introduction} 

The standard model (SM) of electroweak interactions, by far, provides the meticulous portraiture of fundamental interactions of nature. For all its success, SM can not be considered an impeccable theory. This is because of numerous reasons, for e.g., the particle spectrum of SM is deprived of any dark matter candidates. Further, the CKM mechanism fails to account for the observed matter anti-matter asymmetry of the universe. Moreover, there is a large number of unpredicted parameters in the SM which are inserted by hand. Therefore, one needs to look for the signatures of possible physics beyond the SM. This can be accomplished in multifarious ways. Flavor physics serves as a fecund tool to hunt for new physics. It capitalizes on the potential of possible heavy new particles to show footprints on decay modes of hadrons and leptons through quantum loops. These decays are accessible at high luminosity low-energy colliders in contrast to the high energy frontier experiments such as ATLAS and CMS. The decays of B mesons are particularly significant because they offer a wealth of observables to probe new physics, including various 
CP violating observables \cite{Fleischer:2024bpb,Fleischer:2022klb,SinghChundawat:2022zdf,Gangal:2022ole,Das:2022xjg}. These are crucial for investigating additional weak phases that could explain the observed matter-antimatter asymmetry. Additionally, these decays have the potential to set limits on physics at much finer length scales \cite{Alok:2024amd}.  Further, a number of observables in flavour changing neutral current ($b\to sll$) transition shows deviations with SM predictions, for e.g., branching ratio (BR) of $B_s \to \phi\, \mu^+\,\mu^-$  deviates from the SM prediction at 3.5$\sigma$ level \cite{bsphilhc2,bsphilhc3}, and the measured value of angular observable $P'_5$ in $B \to K^* \, \mu^+\,\mu^-$ in $[4.0-6.0] GeV^2$ bin, at the level of 3$\sigma$ \cite{Kstarlhcb1, Kstarlhcb2,LHCb:2020lmf,sm-angular}. \textcolor{black}{These anomalies can be ascribed to new physics (NP) contribution beyond SM in $b\to s\mu^+\mu^-$ process. Nevertheless, the LHCb collaboration has recently revised the measurements for Lepton flavor universality (LFU) ratio ($R_K$ and $R_{K^{*}}$) in $B^0 \to K_S^0 \mu^+  \mu^- $ and $B^+ \to K^{*+} \mu^+ \mu^-$ decay modes ~\cite{LHCb:2022qnv, LHCb:2022vje}. The updated results for $R_K$ and $R_{K^{*}}$ are aligned with SM predictions and provide indications for a universal coupling.}

These aberrant measurements can be analyzed in a model-independent way using the language of effective field theory. There are numerous ways in which such analysis can be done. These can be accommodated by NP in the form of vector and axial-vector 
operators, see for e.g., ~\cite{Alok:2010zd,Alok:2011gv,Descotes-Genon:2013wba,Altmannshofer:2013foa,Hurth:2013ssa,Datta:2019zca,Kumar:2019qbv,Alok:2019ufo,
Altmannshofer:2021qrr,Carvunis:2021jga,Alguero:2021anc,
Geng:2021nhg,Hurth:2021nsi,Angelescu:2021lln,Alok:2022pjb,Ciuchini:2022wbq,SinghChundawat:2022ldm,SinghChundawat:2022zdf,Alok:2023yzg,Alok:2024cyq,Alguero:2023jeh}. For e.g.,  we have several discretions to deal with the hadronic uncertainties and statistical methods. However, it turns out that irrespective of the adopted methodology, the minimal solution requires universal new physics in the form of vector (V) /axial-vector (A) operators. To be more specific, the minimal solution corresponds to a negative value of the Wilson coefficient corresponding to $O_9\equiv(\overline{s} \gamma^{\mu} P_L b)\ (\overline{\mu} \gamma_{\mu}\mu)$ operator, i.e. all allowed solutions include $O_9$ operator.

 New physics in the form of scalar/pseudoscalar (S/P) or tensor (T) operators are not ruled out. However, they can not accommodate current anomalies by itself. In a specific combination with new physics in the form of V/A operators, S/P or T operators can provide a moderate fit to the data \cite{Hurth:2021nsi,231005}. If at all there is new physics in $b \to s \ell \ell$ transition then it should also opine in other decay modes induced by the same quark level transition. There are several such decay modes. 
 
 In this work, we consider one such decay $B\to  K^*_2(1430)(\to K\pi)\ell^+ \ell^-$ where $K^*_2$ is a tensor (parity $2^+$) meson. Like $K^*$, the dominant decay mode of $K^*_2$ is a two-body $K \pi$ mode which is easily accessible at the LHCb. Given the similarity between $B\to  K^* \ell^+ \ell^-$ and $B\to  K^*_2(1430)\ell^+ \ell^-$ decay modes and the fact that $B\to  K^*_2(1430)\gamma$ decay has already been observed by the Belle and BABAR collaborations ~\cite{Belle:2002ekk, BaBar:2003aji}, one can expect measurement of $B\to  K^*_2(1430)\ell^+ \ell^-$ decay mode at the LHCb or Belle-II. 
 The form factor for $B\to K_2^*$ transition, which is non-perturbative, has been computed using various approaches. These includes:  ISGW ~\cite{Isgur:1989}, ISGW2 model~\cite{Scora:1995ty, Sharma:2010yx}, Perturbative QCD ~\cite{Wang:2011}, Light Cone Sum Rule ~\cite{Yang2010qd}, and large energy effective theory (LEET) approach ~\cite{Charles:1990}. Using different form factors $B\to  K^*_2(1430)\ell^+ \ell^-$ decay has been extensively analyzed within SM as well as in various beyond SM scenarios. The scrutiny span across several references including but not limited to ~\cite{Lu:2011jm, Kumbhakar:2022szr, Li:2010ra, Junaid:2011egj, Mohapatra:2021izl,Katirci:2011mt, Ahmed:2012, Das:2018orb,RaiChoudhury:2006bnu,Hatanaka:2010fpr,Aliev:2011gc, Junaid:2011egj}. For instance in ref.~\cite{Li:2010ra} $B\to  K^*_2\mu\mu$ decay has been examined in SM along with vector-like quark model and family non-universal $Z'$ model. Additionally, ref.~\cite{Mohapatra:2021izl} delves into light $Z'$ model considering coupling emerges from $C_9^{NP} i.e.$ vectorial contribution to muon and get remarkable contribution to optimized observable $(P_5')$ and \textcolor{black}{flavor difference observable $(Q_i)$ defined in ref.~\cite{Mohapatra:2021izl} for NP}. Furthermore, the decay $B\to  K^*_2\ell\ell$ has also been investigated with flipped sign of $C_7^{eff}$ and universal extra-dimensional model as in ref.\cite{Katirci:2011mt}. Ref. \cite{Ahmed:2012} investigates the impact of new V/A, S/P, and T type interaction on $B\to  K^*_2\mu\mu$ decay. Recently, LHCb collaboration has provided updated measurements for $R_K$ and $R_K^*$~\cite{LHCb:2022qnv,LHCb:2022vje}, which are consistent with SM predictions. In this work, we exploit the LCSR and LEET to ascertain the transition form factor for $B\to  K_2^*$ transition, and we conduct a comprehensive analysis of $B\to  K^*_2(1430)(\to K\pi)\ell^+ \ell^-$ decay, taking into account new physics in the form of  V/A and S/P  operators, considering scenarios with non- vanishing lepton masses.
 In most of works, the $B\to  K^*_2(1430)(\to K\pi)\ell^+ \ell^-$ decay has been scrutinized with non universal coupling. \textcolor{black}{However, in the presence of universal Wilson coefficients(WCs), universal coupling will induce new physics effects in $b\to s\mu^+\mu^-$ decay.} For various universal NP solutions, we study basic observables in $B\to  K^*_2(1430)(\to K\pi)\mu^+ \mu^-$ decay modes such as the branching ratio $BR$, muon forward-backward asymmetry $A_{FB}$, $ K^*_2$  longitudinal polarization fraction $F_L$ and optimized observables $(P_i)$.  
 It would be interesting to see whether the current data allows for a sizeable new physics effects in some of $B\to  K^*_2(1430)(\to K\pi)\ell^+ \ell^-$ observables. We investigate whether these observables can discriminate between some of the allowed new physics solutions.
 
 In addition, we study this decay in the $Z'$ model, it has  potential to employ these model-independent solutions at tree level. In model-independent analysis, $b\to sll$ transition gives constraints. However, in $Z'$ $B_s-\Bar{B_s}$ mixing and neutrino trident also give additional constraints. We investigate Wilson coefficients in $Z'$ for different scenarios and predict angular observables for $B\to K_2^*(1430)\mu^+\mu^-$ in $Z'$ model.
 
 The paper is organized as follows. In section ~\ref{sec:II}, we start with  effective Hamiltonian for $b\xrightarrow{} sl^+l^-$ transition for SM and NP. Furthermore, we also present the hadronic matrix element for $B\xrightarrow{} K_2^*$ transition. In section ~\ref{sec:III} we discussed $B\xrightarrow{}K_2^*$ transition form factor using different techniques.  In section ~\ref{sec:IV} we calculate fourfold distribution for $B\xrightarrow{} K_2^*( K_2^*\xrightarrow{}K\pi)ll$ for vector, axial- vector and scalar interaction. In section ~\ref{sec:V} we study the SM and NP predictions for angular observables $B\xrightarrow{}K_2^*(\xrightarrow{}K\pi)ll$ decay and the effect of different form factors on these. Further, we investigate the sensitivity of these observables to discriminate various NP solutions. In section ~\ref{sec: VIII} we provide $Z'$ model-dependent analysis for $B\to K_2^*\mu^+\mu^-$ decay. In section ~\ref{sec: VIII} we provide a summary of our results.
\section{Theoretical Framework} 
\label{sec:II}
The decay $B\rightarrow K^*_2(1430)\mu^+ \mu^-$ is mediated by the $ b\to s\mu^+ \mu^- $ quark level transition. In the SM, the effective Hamiltonian for $ b\to s\mu^+ \mu^- $ transition is given by


\begin{widetext}
\begin{equation}
\begin{split}
\mathcal{H}^{\rm SM} &= - \frac{4 G_F}{\sqrt{2} \pi} V_{ts}^* V_{tb}\; [ \sum_{i=1}^{6} C_i(\mu) \mathcal{O}_i(\mu) + C^{\rm eff}_7 \frac{e}{16 \pi^2} [\overline{s} \sigma_{\mu \nu}(m_s P_L  + m_b P_R)b]F^{\mu \nu}  
+ C^{\rm eff}_9 \frac{\alpha_{em}}{4 \pi}(\overline{s} \gamma^{\mu} P_L b)(\overline{\mu} \gamma_{\mu} \mu)\\ &+ C_{10} \frac{\alpha_{em}}{4 \pi} (\overline{s} \gamma^{\mu} P_L b)(\overline{\mu} \gamma_{\mu} \gamma_5 \mu) ],
\end{split}
\end{equation}
\end{widetext}
where $G_F$ is Fermi constant, $V_{tb}$, $V_{ts}$ are the CKM matrix elements, $C_i$ are Wilson coefficients, $P_{L,R} = (1 \mp \gamma^{5})/2$ are the projection operators and $\mathcal{O}_i(\mu)$ are four fermion operators. The effect of the operators $\mathcal{O}_i,\,i=1-6,\, 8 $  embedded in the redefined effective Wilson coefficients  $C_7(\mu)\rightarrow C^{\rm eff}_7(\mu,\, q^2)$ and $C_9(\mu)\rightarrow C^{eff}_9(\mu,\, q^2)$. 
 These redefined effective Wilson coefficients are described in Appendix \ref{WC}~\cite{Buras:1994dj}.
The SM values of the Wilson coefficients can be found in Ref.~\cite{Ali:1999mm}. In a model-independent approach, we can add possible vector(V), axial-vector(A), scalar(S), and pseudo-scalar(P) new physics contributions to the SM effective Hamiltonian. The NP Hamiltonian can be expressed as
\begin{equation}
\begin{split}
\mathcal{H}^{\rm NP}_{\rm VA} &= -\frac{\alpha_{\rm em} G_F}{\sqrt{2} \pi} V_{ts}^* V_{tb} [ C^{\rm NP}_9 \, (\overline{s} \gamma^{\mu} P_L b)\ (\overline{\mu} \gamma_{\mu}\mu)\\& + C^{\rm NP}_{10} \, (\overline{s} \gamma^{\mu} P_L b)\, (\overline{\mu} \gamma_{\mu} \gamma_5 \mu) +C^{\prime}_9 \, (\overline{s} \gamma^{\mu} P_R b)\, (\overline{\mu} \gamma_{\mu}\mu)\\&  + C^{\prime}_{10} \, (\overline{s} \gamma^{\mu} P_R b) (\overline{\mu} \gamma_{\mu} \gamma_5 \mu)], 
\end{split}
\label{HNP}
\end{equation}
\begin{equation*}
\begin{split}
\mathcal{H}^{\rm NP}_{\rm SP} &= -\frac{\alpha_{\rm em}G_F}{\sqrt{2}\pi}V^*_{ts}V_{tb}
[ C_{S} \, (\overline{s}P_R  b)(\overline{\mu}\mu) + C_{P} \, (\overline{s}P_R  b)(\overline {\mu} \gamma_5\mu)\\
&  + C^{\prime}_{S} \, (\overline{s} P_L b)\,(\overline{\mu}\mu) + C^{\prime}_{P} \, (\overline{s}
P_L b)\,(\overline{\mu} \gamma_5\mu)],   
\end{split}    
\end{equation*}

where $C^{\rm NP}_{9,\, 10}$, $C^{\prime}_{9,\, 10}$,  and $C^{'}_{S,\, P}$ are the NP Wilson coefficients.


\subsection{Hadronic matrix elements}
The $B\rightarrow K^*_2(1430)\ell^+ \ell^-$ decay consists of both short-distance and long-distance physics. The short distance physics is embedded in the Wilson coefficients whereas the long distance contribution can be expressed in terms of hadronic matrix elements. \textcolor{black}{For $B\to K^*_2$ transition, the hadronic matrix elements for V and A currents can be parameterized in terms of four form factors $V(q^2)$ and $A_{0,1,2}(q^2)$.} These can be written as~\cite{Wang:2011,Yang2010qd}
\begin{equation*}
\langle K_2^*(k, \epsilon^*)|\bar s\gamma^{\mu}b|\overline B(p)\rangle
  =-\frac{2V(q^2)}{m_B+m_{K_2^*}}\epsilon^{\mu\nu\rho\sigma} \epsilon^*_{T\nu}  p_{\rho}k_{\sigma},    
\end{equation*}
\begin{equation*}
\begin{split}
 \langle  K_2^*(k,\epsilon^*)|\bar s\gamma^{\mu}\gamma_5 b|\overline B(p)\rangle
   &=2im_{K_2^*} A_0(q^2)\frac{\epsilon^*_{T } \cdot  q }{ q^2}q^{\mu}\\& + i(m_B+m_{K_2^*})A_1(q^2)\\&[ \epsilon^{*\mu}_{T}
    -\frac{\epsilon^*_{T } \cdot  q }{q^2}q^{\mu} ] 
    -iA_2(q^2)\\&\frac{\epsilon^*_{T} \cdot  q }{  m_B+m_{K_2^*} }
     [ (p+k)^{\mu}\\& -\frac{m_B^2-m_{K_2^*}^2}{q^2}q^{\mu} ],
\end{split}
\end{equation*}
\begin{equation*}
\begin{split}
\langle K_2^*(k, \epsilon^*)|\bar sq_{\nu}\sigma^{\mu\nu}b|\overline B(p)\rangle&=-2i T_1\epsilon^{\mu\nu\rho\sigma} \epsilon^*_{T\nu}  p_{\rho}k_{\sigma},    
\end{split}
\end{equation*}
\begin{equation}
\begin{split}
    \langle K_2^*(k, \epsilon^*)|\bar sq_{\nu}\sigma^{\mu\nu}\gamma_5 b|\overline B(p)\rangle&= T_2[(m_B^2-m_{K_2^*}^2)\epsilon_{T_\mu}^* \\&-(\epsilon^*.q)P^\mu]\\ & +T_3 (\epsilon^*.q)[q^\mu-\frac{q^2(p+k)^\mu}{m_B^2-m_{K_2^*}^2}].
\end{split}    
\end{equation}
However, it can be shown that the matrix element of $B\to K^*_2$ transition mediated by scalar current ($\bar{s}b$) vanishes and the psedo-scalar interaction leads to 

\begin{equation}
 \langle K_2^*(k, \epsilon^*)|\bar s\gamma_{5}b|\overline B(p)\rangle =  -\frac{2i m_{K^*_2} A_0(q^2)}{m_b + m_s} (\epsilon^*_{T} \cdot  q),
\end{equation}
where $p$ and $k$ are the momentum of $B$ and $K_2^{*}$ meson, respectively. The polarization $\epsilon^{\mu\nu}(n)$ of tensor meson $K_2^\ast$, which has four momentum $(k_0, 0, 0, \vec{k})$, can be written in terms of the spin-1 polarization vectors~\cite{Li:2010ra, Donnachie2000,Wang:2011}

\begin{eqnarray}
 \epsilon_{\mu\nu}(\pm 2) &=& \epsilon_\mu(\pm 1)\epsilon_\nu(\pm 1), \nonumber\\
 \epsilon_{\mu\nu}(\pm 1) &=& \frac{1}{\sqrt{2}}\Big[\epsilon_\nu(\pm)\epsilon_\nu(0) + \epsilon_\nu(\pm)\epsilon_\mu(0) \Big], \nonumber
 \end{eqnarray}
\begin{equation}
\begin{split}
 \epsilon_{\mu\nu}(0) &= \frac{1}{\sqrt{6}}\Big[\epsilon_\mu(+) \epsilon_\nu(-) + \epsilon_\nu(+) \epsilon_\mu(-) \Big] \\& + \sqrt{\frac{2}{3}}\epsilon_\mu(0)\epsilon_\nu(0),   
\end{split}
\end{equation}

where the spin-1 polarization vectors are defined as ~\cite{Kim2000, Faessler:2002ut}:

\begin{equation}
\epsilon_\mu(0) = \frac{1}{m_{K_2^\ast}}(\vec{k_z},0,0,k_0)\, ,\quad \epsilon_\mu(\pm) = \frac{1}{\sqrt{2}}(0,1,\pm i, 0)\ .
\end{equation}

\textcolor{black}{We are studying the decay mode where we have two leptons in the final state makes the virtual particle to have $n=0,\pm1$ helicity states,  so in this case, the $n=\pm 2$ helicity states of the $K_2^\ast$
is not realized. Therefore, a new polarization vector is introduced which behave very similar to spin 1 polarisation vector.}\cite{Wang:2010tz}:
\begin{equation}
\epsilon_{T\mu}(h) = \frac{\epsilon_{\mu\nu}p^\nu}{m_B}\, .
\end{equation}

 The explicit expressions of polarization vectors are
\begin{equation*}
\begin{split}
\epsilon_{T\mu}(\pm 1) &= \frac{1}{m_B}\frac{1}{\sqrt{2}}\epsilon(0).p\  \epsilon_\mu(\pm)\\ & = \frac{\sqrt{\lambda}}{\sqrt{8}m_B m_{K_2}^\ast} \epsilon_\mu(\pm), 
\end{split}
\end{equation*}
\begin{equation}
\begin{split}
\epsilon_{T\mu}(0) &= \frac{1}{m_B}\sqrt{\frac{2}{3}}\epsilon(0).p \ \epsilon_\mu(0)\\&  = \frac{\sqrt{\lambda}}{\sqrt{6}m_B m_{K_2}^\ast} \epsilon_\mu(0).   
\end{split}
\end{equation}
The virtual gauge boson can have, longitudinal, transverse and time-like, polarization states as

\begin{equation*}
\epsilon^\mu_V(0) = \frac{1}{\sqrt{q^2}}(-|\vec{q_z}|,0,0,-q_0)\, ,\quad \epsilon^\mu_V(\pm) = \frac{1}{\sqrt{2}}(0,1,\pm i, 0)\ 
\end{equation*}
\begin{equation}
 \epsilon^\mu_V(t) = \frac{1}{\sqrt{q^2}}(q_0,0,0,q_z)   
\end{equation}
where $q^\mu=(q_0,0,0,q_z)$ is four momentum of gauge boson.
\section{$B\to K_2^*$ transition form factor}
\label{sec:III}
The hadronic matrix element which is involved in the exclusive B mesons decay comprises  several form factors, which are non-perturbative entities. These are the most significant contributors to the uncertainty in theoretical predictions. 
The hadronic matrix elements for V and A currents, for $B\to K_2^*$  transition, can be parameterized in terms of four form factors, $V(q^2)$ and $A_{0,1,2}(q^2)$. They can be calculated using a variety of techniques, some of which include perturbative lattice QCD~\cite{Wang:2011}, QCD Sum Rules, and the Light Cone Sum Rule(LCSR) ~\cite{Cheng:2010hn, Wang:2010tz, Yang2010qd, Aliev:2019ojc}. The large energy effective theory (LEET) and LCSR have been considered for the form factor determination. Under LCSR three distinct parameters can be used to describe the $q^2$ distribution for the B to T transition \cite{Yang2010qd}:\begin{equation}\label{F.F.exp}
F^{B_qT}(q^2)=\frac{F^{B_qT}(0)}{1-a(q^2/m_{B}^2)+b(q^2/m_B^2)^2}\,.
\end{equation}
 where $F^{B_qT}$ can be $V^{B K_2^*}$, $A^{B K_2^*}_{0,1,2}$ or $T^{B K_2^*}_{1,2,3}$. The form factor at zero recoil is denoted here by $F^{B_qT}\left(0\right)$, a and b are considered to be input parameters, and their respective numeric values are given in  Table\ref{F.F.}. \textcolor{black}{The decay constant of the concerned meson, the mass of the b quark, the Borel mass, and the mass of the strange quark are the sources of error in the form factor.} 
%
Within bounds of bulky quark mass $m_b\to \infty$ and enormus energy  of final hadron $ \left(E\to \infty \right)$, $B \to K_2^*$ interaction can be expanded in small ratio $\frac{\lambda_{QCD}}{E}$ and $\frac{\lambda_{QCD}}{m_b}$. In this limit, energy symmetry is achieved at leading power in $\frac{1}{m_b}$, this symmetry significantly simplifies the heavy to light transition ~\cite{Dugan:1991,Charles:1990} and reduces the uncertainty in angular observables. In the large recoil energy domain, the $B \to K _2^*$ transition form factor can be formulated in two independent universal soft form factors, $\zeta_{\perp}\left(q^2\right)$ and $\zeta_{\parallel}\left(q^2\right)$. All the form factors associated with $B \to K_2^*$ transition can be explored in terms of $\zeta_{\perp}\left(q^2\right)$ and $\zeta_{\parallel}\left(q^2\right)$ as well~\cite{Hatanaka:2010fpr, Lu:2011jm} 
\begin{equation}
\begin{split}
  A_{0}\left(q^2\right)&=\frac{m_{K_2^*}}{|p_{K_2^*}|}\left[\left(1-\frac{m_{K_2^*}^2}{m_B}\right)\zeta_{\parallel}\left(q^2\right)+\frac{m_{K_2^*}}{m_B}\zeta_{\perp}\left(q^2\right)\right],\\
  A_{1}\left(q^2\right)&=\frac{m_{K_2^*}}{|p_{K_2^*}|}\frac{2E_{K_2^*}}{m_B+m_{K_2^*}}\zeta_{\perp}\left(q^2\right), \nonumber\\
\end{split}
\end{equation}
\begin{equation}
\begin{split}
A_{2}\left(q^2\right)&=\frac{m_{K_2^*}}{|p_{K_2^*}|}\left(1-\frac{m_{K_2^*}}{m_B}\right)\left[\zeta_{\perp}\left(q^2\right)-\frac{m_{K_2^*}}{E}\zeta_{\parallel}\left(q^2\right)\right], \nonumber\\    
\end{split}
\end{equation}
\begin{equation}\label{F.F.exp1}
\begin{split}
V\left(q^2\right)&=\frac{m_{K_2^*}}{|p_{K_2^*}|}\left(1-\frac{m_{K_2^*}}
{m_B}\right)\zeta_{\perp}\left(q^2\right)
,\nonumber\\
T_1\left(q^2\right)&=\frac{m_{K_2^*}}{|p_{K_2^*}|}\zeta_{\perp}\left(q^2\right),\nonumber 
\end{split}
\end{equation}
\begin{equation}
    \begin{split}
    T_2\left(q^2\right)&=\frac{m_{K_2^*}}{|p_{K_2^*}|}\zeta_{\perp}\left(q^2\right)\left(1-\frac{q^2}{m_B^2-m^2_{K_2^*}}\right),\nonumber\\
    T_3\left(q^2\right)&=\frac{m_{K_2^*}}{|p_{K_2^*}|}\left[\zeta_{\perp}-\left(1-\frac{m_{K_2^*}^2}{m_B^2}\right)\frac{m_{K_2^*}}{E}\zeta_{\parallel}\left(q^2\right)\right].
     \end{split}
\end{equation}
The $q^2$ reliance of the soft form factors $\zeta_{\perp}\left(q^2\right)$ and $\zeta_{\parallel}\left(q^2\right)$  is given by~\cite{Charles:1990, Hatanaka:2010fpr}, \begin{equation}
\zeta_{\parallel,\perp}\left(q^2\right)=\frac{\zeta_{\parallel,\perp}\left(0\right)}{\left(1-q^2/m_B^2\right)^2}.
\end{equation} 
Here $\zeta_{\parallel,\perp}\left(0\right)$ are form factors at zero momentum transfer.  There are several tools to calculate form factors at zero recoil, The Bauer-Stech-Wirbel (BSW) model is one of the tools that might be employed to calculate form factors at zero recoil ~\cite{Wirbel, Hatanaka:2010fpr}. Here, we applied it to determine the form factor at zero recoil. The values derived with the perturbative QCD technique~\cite{Wang:2011}  using the non-trivial relations observed at the large energy limit have been used for our numerical analysis. These values are $\zeta_{\parallel}(0)=0.26\pm 0.10$ and $\zeta_{\perp}(0)=0.29\pm 0.09$.\\
\section{Fourfold angular distribution}
\label{sec:IV}The differential distribution of four-body decay $B\to K^*_2(\to K\pi)\ell^+\ell^-$ can be parametrized as the function of one kinematic and three angular variables. The kinematic variable is $q^2 = (p-k)^2$, where $p$ and $k$ are  four-momenta of $B$ and $K_2^*$ mesons respectively. The $K_2^*$ is moving in the Z direction in the B rest frame. The angular variables are defined in the lepton rest frame. They are (a) $\theta_{K}$ the angle between $K_2^*$ and $K$ mesons where $K$ meson comes from $K^*_2$ decay, (b) $\theta_{\ell}$ the angle between momenta of $\ell^-$ and $K_2^*$ meson and (c) $\phi$ the angle between $K_2^*$ decay plane and the plane defined by the dilepton momenta. Summing over the spins of final state particles, one can obtain the full decay distribution as ~\cite{Li:2010ra, Yang2021, Rajeev:2020aut}.
\begin{widetext}
 \begin{equation}
    \begin{split}
        \frac{d^4\Gamma}{dq^2d\cos{\theta}_{\ell}d\cos{\theta}_K d\phi} &= \frac{15}{128\pi}[ I_1^c({3\cos^2{\theta_K}-1})^2+ I_1^s 3\sin^2{2\theta_K} 
  + I_2^c({3\cos^2{\theta_K}-1})^2 \cos{2\theta}_{\ell} + I_2^s3 \sin^2{2\theta_K} \cos{2\theta}_{\ell} \\
  & +I_3 3\sin^2{2\theta_K}\sin^2\theta_{\ell}\cos{2\phi}
   +I_4 2\sqrt{3} ({3\cos^2{\theta_K}-1})\sin{2\theta}_K\sin{2\theta}_{\ell}\cos{\phi} 
  \\& +I_5 2\sqrt{3}({3\cos^2{\theta_K}-1})\sin{2\theta}_K\sin{\theta}_{\ell}\cos{\phi} 
    +I_6^s3\sin^2{2\theta}_K\cos{\theta}_{\ell} +I_6^c({3\cos^2{\theta_K}-1})^2\cos{\theta}_{\ell} \\
   & + I_7 2\sqrt{3}({3\cos^2{\theta_K}-1})\sin{2\theta_K}\sin{\theta}_{\ell}\sin{\phi} 
    + I_8 2\sqrt{3}({3\cos^2{\theta_K}-1})\sin{2\theta}_{K}\sin{2\theta}_{\ell}\sin{\phi}  \\
  &  +I_9 3\sin^2{2\theta}_K\sin^2\theta_{\ell} \sin{2\phi} ]. 
  \end{split}
\label{Fourfold}
\end{equation}
\end{widetext}
The angular coefficients $I_i$'s in the equation \ref{Fourfold}  are function of  $q^2$, which can be written in terms of various transversity amplitudes as:
\begin{equation*}
\begin{split}
 I_1^c &= \left(|A_{0L}|^2+|A_{0R}|^2\right)+\frac{8m_{\ell}^2}{q^2}{\rm Re}\left[A_{0L}A_{0R}^*\right]+4\frac{m_{\ell}^2}{q^2}|A_t|^2\\&+(\beta_{\ell}^2 |A_S|^2),   
\end{split}
\end{equation*}
\begin{eqnarray}
\begin{split}
 I_1^s &=\frac{3}{4}\left(1-\frac{4m_{\ell}^2}{3q^2}\right)\left[|A_{\perp L}|^2+|A_{\parallel L}|^2+ L \Leftrightarrow R \right]\\&+ 
 \frac{4m_{\ell}^2}{q^2} {\rm Re} \left[A_{\perp L}A_{\perp R}^*+A_{\parallel L}A_{\parallel R}^*\right],\\ 
 I_2^c &=  -\beta_{\ell}^2\left[|A_{0L}|^2+|A_{0R}|^2\right],\\
 I_2^s &= \frac{\beta_{\ell}^2}{4} \left[|A_{\perp L}|^2+|A_{\parallel L}|^2 +L \Leftrightarrow R \right],\\
 I_3 &= \frac{\beta^2_{\ell}}{2}\left[\left(|A_{\perp L}|^2-|A_{\parallel L}|^2 + L \Leftrightarrow R \right) \right],\\
 I_4 &= \frac{\beta_{\ell}^2}{\sqrt{2}}{\rm Re} \left[A_{0L}A_{\parallel L}^*+ A_{0R}A_{\parallel R}^*\right],\\
 I_5&=\sqrt{2}\beta_{\ell}{\rm Re}\Bigg[A_{0L}A_{\perp L}^*-  A_{0R}A_{\perp R}^*\Bigg],\\
 I_6^s &=  -2\beta_{\ell} {\rm Re}\left[A_{\parallel L}A_{\perp L}^*- A_{\parallel R}A_{\perp R}^*\right],\\
 I_6^c &=  4\beta_{\ell}{\rm Re}\left[\frac{m_{\ell}}{\sqrt{q^2}}\left(A_{0L}+A_{0R}\right)A_s^*\right],\\
 I_7 &=-\sqrt{2}\beta_{\ell} {\rm Im}\Bigg[A_{0L}A_{\parallel L}^*-A_{0R}A_{\parallel R}^*\\ & +\frac{m_{\ell}}{\sqrt{q^2}} \left( A_{\perp L}^*+A_{\perp R}^*\right)A_S\Bigg],\\
 I_8 &= \frac{\beta_{\ell}^2}{\sqrt{2}}{\rm Im}\left[\left(A_{0L}A_{\perp L}^*+ A_{0R}A_{\perp R}^* \right)\right],\\
 I_9 &=\beta_{\ell}^2 {\rm Im}\Bigg[\left(A_{\parallel L}^*A_{\perp L} + A_{\parallel R}^*A_{\perp R}\right) \Bigg],
\end{split}
\end{eqnarray}\label{Angular eqn}
 where $\beta_l = \sqrt{1-4m_l^2/q^2}$.


From the angular distribution, one can derive observables such as the forward-backward asymmetry  $A_{\rm FB}$ and the differential decay width $d\Gamma/dq^2$ as a function of the dilepton invariant mass $q^2$ as defined in~\cite{Li:2010ra,Yang2021,Mohapatra:2021izl}.
\begin{itemize}
	\item  The differential decay width $d\Gamma/dq^2$  can be determined by integrating over the angles $\theta_l,\theta_k,$ and, $ \phi $ as \cite{Li:2010ra,Yang2021}:
	\begin{eqnarray}\label{eq:diffdecaywidth}
	\frac{d\Gamma}{dq^2} &=& \frac{1}{4}(3I_1^c + 6I_1^s - I_2^c - 2I_2^s)\, .
	\end{eqnarray}
	\item The forward-backward asymmetry of lepton pair $A_{\rm FB}$ (normalized by differential decay width) is given by \cite{Li:2010ra,Yang2021},
	\begin{eqnarray}\label{eq:AFB}
	A_{\rm FB}(q^2) &=& [\int_{0}^{1}-\int_{-1}^{0}]d\cos{\theta}\frac{d^2\Gamma}{dq^2d\cos{\theta_l}}\,,
	\end{eqnarray}
	\begin{eqnarray}\label{eq:AFB}
	\Bar{\frac{dA_{\rm FB}}{dq^2}} &=& \frac{3I_6}{3I_1^c + 6I_1^s - I_2^c - 2I_2^s}\,.
	\end{eqnarray}
 
	\item Similarly fourfold distribution for CP conjugate process $\bar{B}\to \bar{K_2^*}(\bar{K_2^*}\to K\pi)l^+l^-$ is given as:
	\begin{equation*}
	\frac{d^4\Gamma}{dq^2d\cos{\theta}_{\ell}d\cos{\theta}_K d\phi} = \frac{15}{128\pi}\sum_{i=1}^{9}\bar{I_i}(q^2,\theta_l,\theta_k,\phi)
	\end{equation*}
	Since CP transformation interchange lepton and antilepton, leads to a transformation $\theta_l \to \theta_l-\pi$, and $\phi \to -\phi$ This leads to  $I_{1,2,3,4,7}^{(a)}\to \bar{I}_{1,2,3,4,7}^{(a)}$, $I_{5,6,8,9}^{(a)}\to -\bar{I}^{(a)}_{5,6,8,9}$. One can construct optimized observables with reduced uncertainty as defined in Refs. ~\cite{Descotes_Genon_2013, Descotes-Genon:2013vna} :

 \begin{equation*}
	<P_1>=\frac{1}{2}\frac{\int dq^2 I_3}{\int dq^2 I_{2}^s},\quad  <P_2> = \frac{1}{8}\frac{\int dq^2 I_6^s}{\int dq^2I_{2}^s},
	\end{equation*}
 \begin{equation}
 \begin{split}
 <P_4'> & =\frac{\int dq^2I_4}{\sqrt{-\int dq^2 I_2^s \int dq^2 I_2^c}},\\
<P_5'> &=\frac{\int dq^2 I_5}{2\sqrt{-\int dq^2I_2^s \int dq^2 I_2^c}} 
\end{split}
\end{equation}
 \end{itemize}


\section{Analysis Of $B\to K_2^*(1430)(\to K\pi)l^+l^-$: A Model Independent Approach}
\label{sec:V}
In this segment, prediction for a number of angular observables is done for $B\to  K_2^*(1430)(\to K\pi)l^+l^-$ decay for low $q^2$ within SM and different NP scenarios. In the expression of decay rate and other angular observables, the scalar and pseudoscalar amplitudes are suppressed by lepton mass as in angular coefficients $I_i$. These operators are heavily constrained from $B_s\to \mu^{\pm} \mu^{\mp}$ decay ~\cite{010428, Arbey:2018ics}. \textcolor{black}{In our analysis, we are considering $B\to K_2^*(1430)(\to K\pi)\mu^+\mu^-$ for the smallness of the mass of $m_l=m_\mu$, the contribution of the scalar operators are negligible. So, we analyzed $B\to K_2^*(1430)(\to K\pi)\mu^+\mu^-$ decay without taking them.}
We focus on NP scenarios which comprises of either one or two non-zero Wilson Coefficients. Choosing low $q^2$ region as $1 \leq  q^2\leq 6$  GeV$^2$, we provide angular observable predictions for the $B\to K_2^*(1430)(\to 
 K\pi)\mu^+\mu^-$ decay by employing two distinct sets of form factors for $B\xrightarrow{} K_2^*$ transition. The relevant numeric expressions of form factors are given in eqn. $[\ref{F.F.exp},\ref{F.F.exp1}]$, Table \ref{F.F.} provides the estimated value of fit parameters for these form factors. Our goal is to scrutinize observables for $B\to K_2^*(1430)(\to K\pi)\mu^+\mu^-$ within various New Physics scenarios and look for deviation from SM predictions.
\par Considering recent flavour anomalies like $B_s\rightarrow \mu^+\mu^-$\cite{Combination,LHCb:2021awg,ATLAS:2018cur,CMS:2019bbr,LHCb:2017rmj,Altmannshofer:2021qrr}, $B_s\rightarrow \phi\mu^+\mu^-$ \cite{bsphilhc2,bsphilhc3}, the global fit was performed by several groups to all $b\rightarrow s\mu^+\mu^-$ data for determining the pattern of Wilson coefficients which provides a good fit to data. These fit suggest that several scenarios can mitigate tension between experimental data and theory. We consider three scenarios, considering one operator or two related operators at a time. These correspond to vectorial $C_9^{NP},\, C_9'^{NP}$, and axial vectorial $C_{10}^{NP}$ contribution to muon.
Before December 2022 updates, real fit for $b\xrightarrow{} s\mu^+\mu^-$ shows that $C_9^{NP}$, and $C_9^{NP}= -C_{10}^{NP}$ provide a good fit to data. Whereas $C_9^{NP} = -C_9'^{NP}$ provides a moderate fit \cite{Alok:2022pjb}. In ref \cite{Altmannshofer:2021qrr}, it was shown that along with these, a moderate solution $C_{10}^{NP}$ can also explain B anomalies.
\begin{table}[h!]

    \begin{tabular}{|c | c| c |}
    \hline
    \hline
     Solutions & Wilson coefficients & 1$\sigma$ range\\
    \hline
    S1 & $C_9^{NP}$ &  $-1.08\pm 0.18 $  \\
    \hline
    S2 & $C_9^{NP}= -C_{10}^{NP}$ & $-0.50\pm 0.12$  \\
    \hline
    S3 & $C_9^{NP}= -C_{9}^{'NP}$ & $-0.88\pm 0.16$  \\
    \hline
    \end{tabular}
\caption{Universal NP scenarios suggested as an explanation for all $b\xrightarrow{} sl^+l^-$ anomalies. The numeric value of the Wilson coefficients is taken from\cite{Alok:2023yzg}.}
    \label{WC}
\end{table}
\par However, updated fit after considering updated  measurement for $R_K,$ $R_K^*$ by LHCb December 2022 \cite{LHCb:2022qnv} \cite{LHCb:2022vje} alongside the latest measurement of branching ratio $B_s\xrightarrow{} \mu^+\mu^-$ \cite{BPH} and several observables  for $B_s\xrightarrow{} \phi\mu^+\mu^-$ by CMS\cite{bsphilhc3} suggest that non universal coupling $C^V_{9/10}$  i.e. $b\xrightarrow{} s\mu^+\mu^-$ is disfavored \cite{Alok:2023yzg}. \textcolor{black}{The alignment of the updated LHCb measurements $R_K$, $R_K^*$(LHCb) with Standard Model predictions indicates a potential preference for universal coupling in $b\to sll$ transitions.} Results after considering universal Wilson Coefficients still supports $C_9^{NP}$ is the best scenario, and $C_9^{NP} = -C_{9}'^{NP}$ , $C_9^{NP} = -C_{10}^{NP}$ to be a good solution as shown in Table\ref{WC}.\par Further, we anticipate $q^2$ dependency of the angular observables using all input parameters, including $V_{CKM}$ matrix element and Wilson coefficients that are listed in Table \ref{input parameter} and Ref. \cite{Ali:1999mm} respectively. Table \ref{LEFT+LEET} shows the observable central value and corresponding $1\sigma$ uncertainties for all observables like differential branching ratio, the normalized forward-backward asymmetry $A_{FB}$, Longitudinal polarization fraction $F_L$, and for optimized observables such as $P'_{4,5}$ and  $P_{1,2}$ for $1 \leq  q^2\leq$ 6 GeV$^2$. The third column comprised of the integrated observables using  LCSR form factor and $B\to K_2^*$ transition at large recoil within SM. The result of different choices of form factors have a considerable impact on the angular observables, this can be seen from Table \ref{LEFT+LEET}, where the angular observables have been predicted for SM and different NP scenarios at $\mu =4.2$ GeV. \textcolor{black}{In appendix Fig.\ref{fig_SM} suggests that within standard model branching ratio (BR) for $B\to K_2^*(\to K\pi)\ell^+\ell^-$ $(\ell = \mu)$ decay with LCSR form factor have a significant $q^2$ deviation from BR using LEET form factor formulation.}  In Fig\ref{fig2}, we present $q^2$ dependency for these observables in low dilepton mass region, [0.045, 6.0] GeV$^2$ for SM and different NP scenarios following LCSR form factor. 
\newline
\begin{figure}[t!]
\centering
\includegraphics[width=0.5\textwidth]{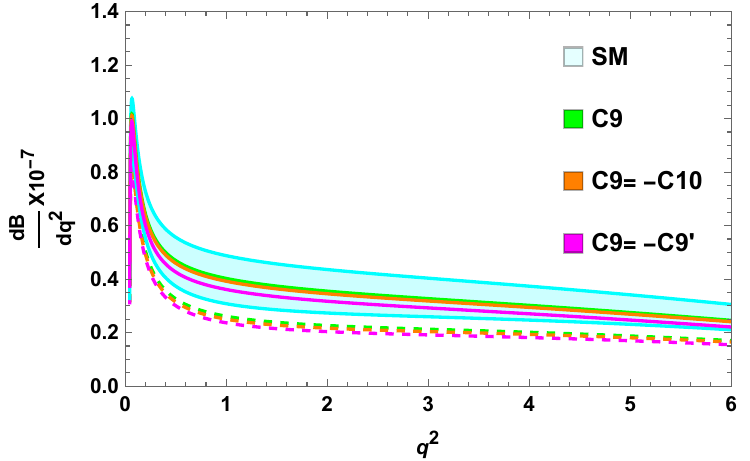}
\caption{Plot for $q^2$ distribution in SM as well as for several new physics scenarios for branching ratio of $B\to K_{2}^{*}(\to K\pi)\mu^+\mu^-$ decay. The standard model is denoted with the cyan band. The solid and dotted lines represent the maximum and minimum boundaries for each new physics scenario.}
\label{fig1}
\end{figure}
\begin{figure*}[ht!]
\centering
\includegraphics[width=85mm]{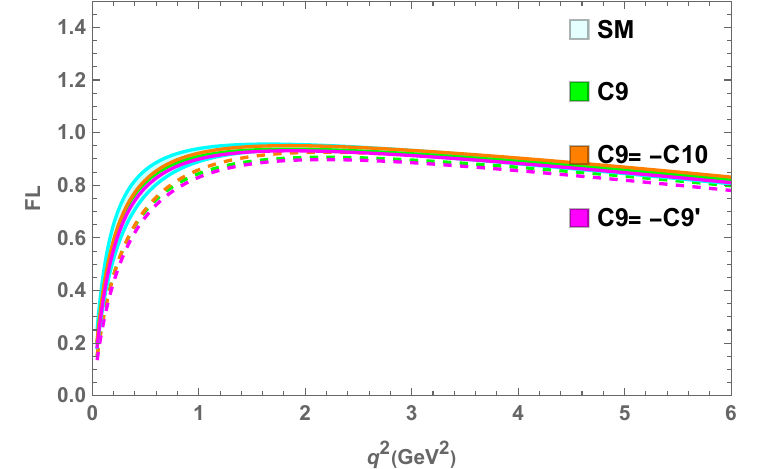}
\includegraphics[width=85mm]{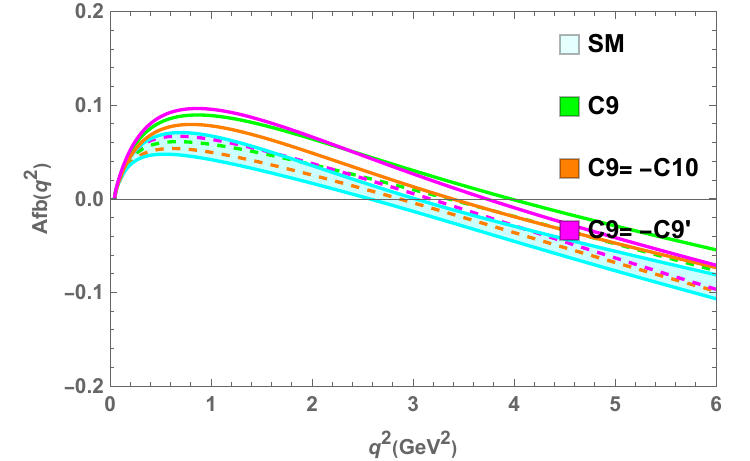}

\caption{The figure delineate $q^2$ dependency for Longitudinal Polarization $F_L$ and forward-backward asymmetry ($A_{FB}$) for $B \to K_{2}^{*}(\to K\pi)\mu^+\mu^-$
 decay Within the confines of the Standard Model and across various new physics scenarios. The cyan band for Standard Model one sigma region. The solid and dotted lines represent the maximum and minimum boundaries for each new physics scenario.}
\label{fig2}
\end{figure*}
\begin{figure*}[ht!]
\centering
\includegraphics[width=85mm]{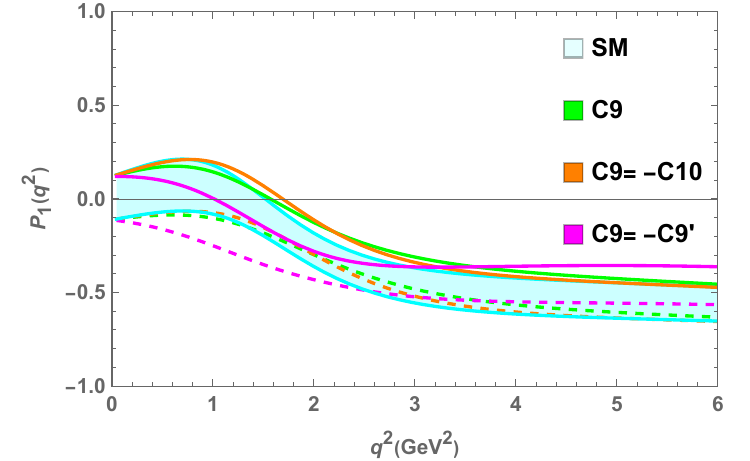}
\includegraphics[width=85mm]{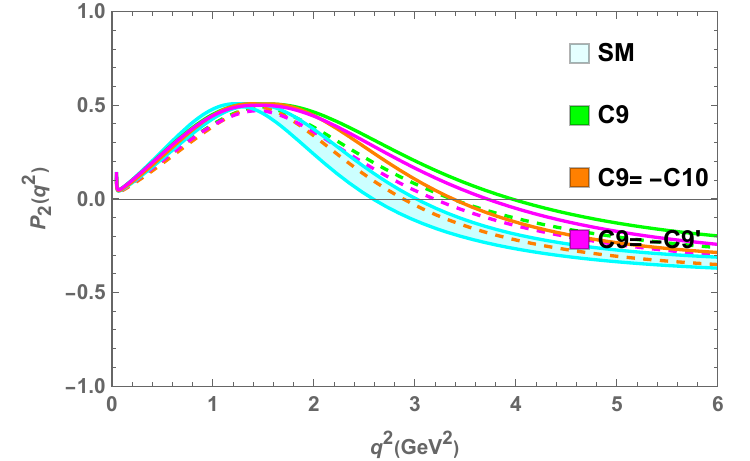}
\\
\includegraphics[width=85mm]{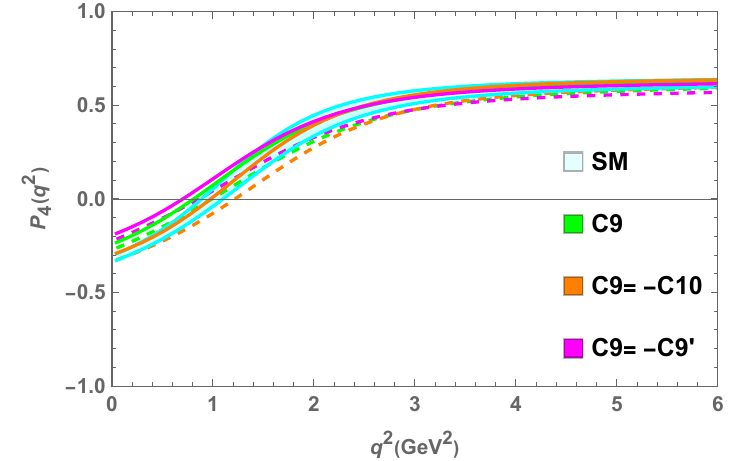}
\includegraphics[width=85mm]{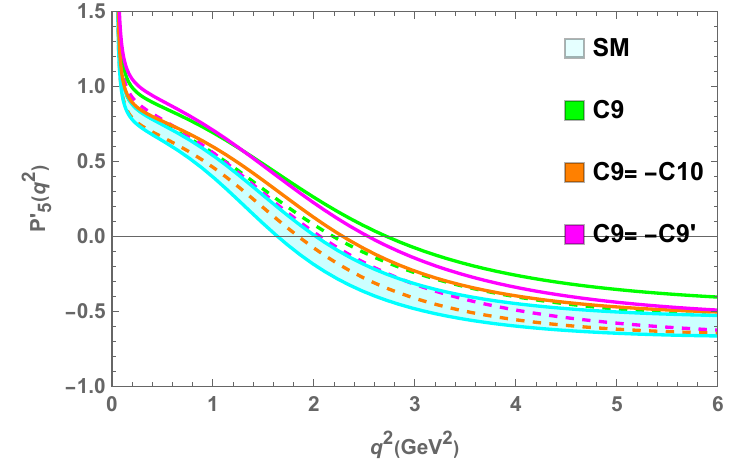}
\\
\caption{Plots for $q^2$ distribution in SM as well as for several new physics scenarios for optimized angular observables $P_1$, $P_4'$ left pannel and $P_2$, $P_5'$ right pannel for $B\to K_2^*\mu^+\mu^-$ decay. The standard model is filled with the cyan band. The solid and dotted lines represent the maximum and minimum boundaries for each new physics scenario.}
\label{fig3}
\end{figure*}
The branching ratio BR, angular observables $F_L$, $A_{FB}$ and optimized angular observables $P_{1,2}$, $P'_{4,5}$  computed for the three updated NP scenarios in low $q^2 \, \epsilon \, [1,6]$ GeV$^2$ as listed in Table \ref{LEFT+LEET} and compared with SM for LCSR formulation in low $q^2$. \textcolor{black}{ In Fig\ref{fig1}, we show the behavior of branching ratio with $q^2$ for SM and NP scenarios with LCSR form factor.} This is also evident from Fig.\ref{fig1} that after updated measurements of $R_K$, $R_{K^*} $ there is suppression in branching ratio for universal Wilson Coefficients in $b\to s\ell^+\ell^-(\ell=e,\mu)$. It is apparent from Table \ref{LEFT+LEET} that the branching ratio for $B\to K_2^*(1430)\mu^+\mu^-$ is the order of $~O(10^{-7})$ in SM as well as for all the NP scenarios. For the S1 scenario, there is a suppression in the value of branching ratio of nearly about $\sim$15$\%$. Furthermore, scenarios S2 and S3  lead to a larger suppression up to $25\%$ from the SM value as compared to scenario S1. 

\begin{table*}[tb!]
\centering
\begin{tabular}{ | m{5em} | m{1.5cm}| m{2.5cm}| m{2.5cm} |m{2.5cm}| m{2.5cm}| } 
 \hline
  Angular Observable & Form factor& SM & S1 &  S2  & S3 \\ 
  \hline
  BR$\times 10^7$& LCSR LEET & $1.554\pm0.558$ $1.212\pm 0.605$ & $1.2696\pm0.446$ $1.041\pm0.515$  & $1.2204\pm0.446$  $0.967\pm 0.484$ &  $1.136\pm0.392$ $0.959\pm0.471$ \\
  \hline
  $F_{L}$ & LCSR LEET & $0.835\pm 0.016$ $0.721\pm 0.105$ & $0.821\pm0.021$ $0.689\pm0.116$  & $0.8234\pm0.0193$ $0.715\pm0.108$ &$0.811\pm0.023$ $0.667\pm0.123$\\
  \hline
  $A_{FB}$ & LCSR LEET & $-0.015\pm 0.011$  $-0.050\pm 0.026$& $0.0087\pm0.015$ $0.0\pm0.012$ & $-0.0053\pm0.013$ $-0.028\pm0.015$ & $0.00419\pm 0.015$  $-0.007\pm0.012$\\
  \hline
  $P_{1}$ & LCSR LEET& $-0.446\pm 0.149$  $0.374\pm 0.001$ & $-0.389\pm0.142$ $-0.008\pm0.065$ & $-0.414\pm0.146$  $0.219\pm 0.047$& $-0.406\pm0.135$ $-0.003\pm0.060$\\
  \hline
  $P_{2}$ & LCSR LEET & $-0.112\pm 0.086$  $-0.171\pm 0.001$& $0.056\pm 0.089$ $0.018\pm0.032$ & $-0.039\pm0.096$ $-0.093\pm0.023$& $0.0247\pm0.092$ $-0.018\pm0.028$\\
  \hline
 $P'_{4}$ &LCSR LEET & $0.975\pm 0.111$  $0.879\pm 0.010$ & $0.934\pm0.114$  $0.840\pm0.011$ & $0.921\pm0.125$ $0.823\pm 0.0161$& $0.932\pm0.104$  $0.832\pm0.012$\\
 \hline
 $P'_{5}$ & LCSR LEET & $-0.349\pm 0.137$  $-0.548\pm 0.0032$ & $-0.138\pm 0.139$ $-0.302\pm 0.049$ & $-0.274\pm 0.146$ $-0.470\pm 0.027$& $-0.201\pm 0.151$ $-0.383\pm 0.045$\\
 \hline
\end{tabular}
\caption{Angular Observables SM and NP predictions with 1$\sigma$ for $q^2$ [1-6] GeV$^2$ bin following LCSR and LEET $B\xrightarrow{}K_2^*$ transition form factors \cite{Yang2010qd, Das:2018orb}, for different New Physics scenarios.\cite{Alok:2023yzg}}
 \label{LEFT+LEET}
\end{table*}    


\begin{table}[hb!]
\centering
    \begin{tabular}{ | c ||c| c|c| }
    \hline
    \hline
     Observable & Solutions &\hspace{1.5cm} $q^2$ Range & \hspace{1.0cm}Predictions\\
         \hline
   & SM & $[q_{min}^2-2.8]
    $ & $0.034\pm 0.012 $  \\
     & & $[2.8-q_{max}^2]
    $ &  $-0.043\pm 0.014 $  \\
    &S1 & $[q_{min}^2-3.65]
    $ &  $0.043\pm 0.015 $  \\
   &  & $[3.65-q_{max}^2]
    $ &  $-0.030\pm 0.015 $  \\
   $A_{FB}$ & S2 & $[q_{min}^2-3.13]
    $ &  $0.038\pm 0.013 $  \\
    & & $[3.13-q_{max}^2]
    $ &  $-0.039\pm 0.015 $  \\
     & S3 & $[q_{min}^2-3.44]
    $ &  $0.047\pm 0.015 $  \\
   &  & $[3.44-q_{max}^2]
    $ &  $-0.038\pm 0.017 $  \\
    \hline
  
   & SM & $[q_{min}^2-1.8]
    $ &  $0.511\pm 0.099 $
    \\
     & & $[1.8-q_{max}^2]
    $ &  $-0.458\pm 0.133 $  \\
   & S1 &$[q_{min}^2-2.42]
    $ &  $0.545\pm 0.108 $  \\
    & & $[2.42-q_{max}^2]
    $ &  $-0.323\pm 0.127 $  \\
    $P_5'$& S2 & $[q_{min}^2-2.04]
    $ &  $0.515\pm 0.099 $  \\
    &  & $[2.04-q_{max}^2]
    $ &  $-0.425\pm 0.139 $  \\
    & S3 & $[q_{min}^2-2.29]
    $ &  $0.578\pm 0.117 $  \\
     & & $[2.29-q_{max}^2]
     $ &  $-0.390\pm 0.138 $  \\
     \hline
     & SM & $[q_{min}^2-2.81]
    $ &  $0.213\pm 0.008 $  \\
     & & $[2.81-q_{max}^2]
    $ &  $-0.258\pm 0.065 $  \\
   & S1 & $[q_{min}^2-3.67]
    $ &  $0.094\pm 0.053 $  \\
    & & $[3.67-q_{max}^2]
    $ &  $0.222\pm 0.015 $  \\
    $P_2$& S2 & $[q_{min}^2-3.15]
    $ &  $0.206\pm 0.009 $  \\
    &  & $[3.15-q_{max}^2]
    $ &  $-0.230\pm 0.072 $  \\
    & S3 & $[q_{min}^2-3.47]
    $ &  $0.218\pm 0.015 $  \\
     & & $[3.47-q_{max}^2]
     $ &  $-0.180\pm 0.072 $  \\
     \hline
    \end{tabular}
\caption{Predictions for forward-backward asymmetry ($A_{FB}$) and optimized observable$P'_5$ and $P_2$, on both side of their zero crossing value for  $B \to K_{2}^{*}(\to K\pi)\mu^+\mu^-$
 decay, here $q_{min}^2=4m_{\mu}^2$ and $q_{max}^2=6$ $GeV^2$.}
    \label{P5}
\end{table} 
 Forward-backward asymmetry ($A_{FB}$) has a zero crossing for SM as well as for NP. Furthermore, in the presence of NP it shifted towards high $q^2$.  $A_{FB}$ is positive for $q^2 <q_{0}^{2}$ and negative for $q^2>q_0^2$ as shown in Fig.\ref{fig2}, where $q_0$ is corresponding value of $q^2$ at zero crossing point. We also give $A_{FB}$ predictions for  $q^2 <q_0^2$ as well as for $q^2 > q_0^2$ as shown in Table\ref{P5}.
From the right of Fig\ref{fig2}, it is clear that new physics can enhance forward-backward asymmetry $A_{FB}$ of muon from SM predictions. For all NP scenarios, the $K_2^*$ longitudinal polarization $(F_L)$ for the low di-lepton mass square, renders the values close to its SM value as can be seen from Fig\ref{fig2}. 
The optimized angular observable $P_1$ gives an enhancement about $10\% - 15\%$ in all the NP scenarios for low $q^2$ bin by SM predictions. S1 and S3 speculate positive values for $P_2$, S2 scenario suggests a negative value for $P_2$. Whereas SM suggests comparatively larger negative values. Therefore, precise measurement of $q^2$ distribution of $P_2$ within a low $q^2$ region could serve to distinguish S2 from the other two scenarios.  This can also be seen from Fig.\ref{fig3}. In the presence of NP zero crossing point for $P_2$ shifts towards higher $q^2$ as can be seen from Fig.\ref{fig3}. The predictions for the optimized angular observable $P_2$ are shown in the top right panel of Fig\ref{fig3} shows enhancement from SM predictions for low $q^2$. However, there is no meaningful deviation from the SM in the observable $P_4'$ for all the NP scenarios under consideration. This can be seen in Fig \ref{fig3}, that the maximum and minimum possible values of observable $P_4'$ in all the NP scenarios overlap with the SM band shown in color cyan. The optimized observable $P_5'$ shows enhancement from the corresponding SM values in all universal NP scenarios, and the zero crossing point shifts towards high $q^2$ in the presence of NP, S2 scenario gives an enhancement of about $22\%$ whereas in S3 scenario enhancement is almost $42\%$ in low $q^2$ bin. Nevertheless, S1 scenario shows nearly two-fold enhancement from SM predictions in the low $q^2$ bin which makes $P_5'$ a valuable observable to study NP effects. From the bottom right panel of Fig \ref{fig3} it can be observed that NP enhance optimized observable $P_5'$ from its SM prediction for 1 GeV$^2$ $\leq  q^2\leq 6$ GeV$^2$. Similar to $A_{FB}$, we calclated  $P_2$ and $P_5'$ for $q^2< q_0^2$ and $q_0^2 > q^2$ as depicted in Table\ref{P5}.


\section{$Z'$ Model}
In this section, we study $B\rightarrow K^*_2(1430)\mu^+ \mu^-$ decay in a non-universal $Z'$ model, where $Z'$ adheres additional $U(1)'$ symmetry. In the presence of $Z'$, the FCNC process $b\to s\mu^+\mu^-$  accomplished at tree level. It couples to left and right-handed muons exclusively without interacting with leptons belonging to other generations. Furthermore, it couples to both left and right-handed quarks. Flavor-changing interaction for $b\to s\mu^+\mu^-$ transition in $Z'$ model is given by \cite{Crivellin:2015era}:
\begin{eqnarray}
L_{Z'} & \supset& J^\alpha Z'_\alpha \nonumber \\   
J^\alpha &  \supset & g_L^{bs}\bar s\gamma^\alpha P_L b\, + \,g_L^{\mu\mu}\bar \mu\gamma^\alpha P_L \mu + h.c.
\end{eqnarray}
Here $g_{L,R}^{\mu\mu}$ and $g_{L,R}^{bs}$ are left and right-handed coupling of $Z'$ boson with muons and quarks respectively. After integrating out heavy $Z'$ we get an effective four-fermion Hamiltonian which not only induces $b\to s \mu \mu$ transition but also generates $B_s -\bar B_s$ mixing. The relevant effective Hamiltonian can be written as \cite{Alok:2021pdh,Alok:2019xub}:
\begin{widetext}
\begin{equation}
\begin{split}
\mathcal{H}_{eff}^{Z'}\supset & \frac{g_L^{bs}}{M_{Z'}^2}\,(\bar s\gamma^\alpha P_L b)\,[\bar \mu\gamma_\alpha \,(g_L^{\mu\mu}P_L+ g_R^{\mu\mu}P_R)\,\mu]\,  + \frac{g_R^{bs}}{M_{Z'}^2}\,(\bar s\gamma^\alpha P_R b)\,[\bar \mu\gamma_\alpha (g_L^{\mu\mu}P_L+ g_R^{\mu\mu}P_R)\,\mu]\\& +    \frac{(g_L^{bs})^2}{2M_{Z'}^2}\,(\bar s\gamma^\alpha P_L b)\,(\bar s\gamma_\alpha P_L b)\, +\frac{(g_R^{bs})^2}{2M_{Z'}^2}\,(\bar s\gamma^\alpha P_R b)\,(\bar s\gamma_\alpha P_R b)\\&+\frac{(g_L^{bs}g_R^{bs})}{M_{Z'}^2}(\bar s\gamma^\alpha P_L b)\,(\bar s\gamma_\alpha P_R b).
\end{split}
\label{HZ}
\end{equation}    
\end{widetext}
In Eqn. \ref{HZ} initial two terms correspond to $b\to s\mu^+\mu^-$ transition and the rest terms induce $B_s -\bar B_s$ mixing. In the presence of $Z'$  the Wilson coefficients in $b\to s\mu^+\mu^-$ transition get modified as:
\begin{eqnarray}
    C_{9, 10}&=& C_{9, 10}^{SM}+ C_{9, 10}^{NP}\nonumber\\
 \label{WCSZ}
\end{eqnarray}
after matching with eqn. \ref{HNP} we get,
\begin{eqnarray}
     C_{9}^{NP}&=&\frac{-N_1}{2}g_{L}^{bs}(g_L^{\mu\mu}+g_R^{\mu\mu}) \nonumber\\
     C_{10}^{NP}&=&\frac{N_1}{2}g_{L}^{bs}(g_L^{\mu\mu}-g_R^{\mu\mu}).
     \label{C9}
\end{eqnarray}

The right-handed quark coupling influences chirally flipped Wilson coefficients $C_{9,10}^{'}$, as:
\begin{eqnarray}
    C_9^{'}&=&\frac{-N_1}{2}g_{R}^{bs}(g_L^{\mu\mu}+g_R^{\mu\mu})\nonumber\\
     C_{10}^{'}&=&\frac{N_1}{2}g_{R}^{bs}(g_L^{\mu\mu}-g_R^{\mu\mu}).    
\end{eqnarray}
Where $N_{1}=\sqrt{2}\pi/(\alpha_{em}V_{tb}V_{ts}^{*}M_{Z^{'}}^{2})$.
It is evident from universal $b\to sll$ fitting that $C_9^{NP}<0$, $C_9^{NP}=-C_{10}^{NP}$ and $C_9^{NP}=-C_9^{'NP}$ generates a good fit to data \cite{Alok:2023yzg}.
It can be seen from Eqn. \ref{C9} that right handed quark coupling $(g_R^{bs})$ do not contribute in $C_{9}^{NP}$ and $C_{10}^{NP}$. S1 NP scenario can be achieved by substituting $g_L^{\mu\mu}=g_R^{\mu\mu}$ along with $g_R^{bs}=0$. While the S2 NP scenario can be obtained using $g_R^{\mu\mu}=$ $g_R^{bs}=0$. However, the  S3 NP scenario can be induced by using  $g_L^{\mu\mu}=g_R^{\mu\mu}$,
and  $g_L^{bs}=g_R^{bs}$. We now proceed to discuss $B_s -\bar B_s$ mixing.
The NP effective Hamiltonian for $B_s- \bar B_s$ mixing can be described as: 
\begin{eqnarray}
 \mathcal{H}^{\Delta B=2} & \supset & \frac{4G_f}{\sqrt{2}}(V_{tb}V_{ts}^*)\,[C_1^{bs}\,(\bar s\gamma^\alpha P_Lb)^2+ C_2^{bs} (\bar s\gamma^\alpha P_Rb)^2 \nonumber\\ &  & + C_3^{bs}\,(\bar s\gamma^\alpha P_Lb)\,(\bar s\gamma^\alpha P_Rb)\,].
]
\label{HBsZ}    
\end{eqnarray}
after matching with eqn.\ref{HZ} NP Wilson coefficients are given as:
\begin{eqnarray}
C_1^{bs}&= &\frac{1}{4\sqrt{2}G_f M_{z'}^2}\left(\frac{g_L^{bs}}{V_{tb}V_{ts}^*}\right)^2 \nonumber\\ 
C_2^{bs}&=& \frac{1}{4\sqrt{2}G_f M_{z'}^2}\left(\frac{g_R^{bs}}{V_{tb}V_{ts}^*}\right)^2 \nonumber\\ 
C_3^{bs}&=& \frac{1}{4\sqrt{2}G_f M_{z'}^2}\left(\frac{g_L^{bs}g_R^{bs}}{(V_{tb}V_{ts}^*)^2}\right).  
\label{BsBs}
\end{eqnarray}
Neglecting scalar and tensor operators, which are currently disfavored by $b\xrightarrow{}s$ data, the contribution to $B_s -\bar B_s$ mixing normalized to the Standard Model is represented by 
\begin{widetext}
\begin{equation}
\begin{split}
\frac{\Delta M_s^{SM+NP}}{\Delta M_s}&=   \Bigg| 1+\frac{\eta^{6/23}}{R_{loop}^{SM}}\, \biggr[C_1^{bs}+C_2^{bs}\\ & -\frac{C_3}{2\eta^{3/23}}\left(\frac{B_5}{B_1}\left( \frac{M_{B_s}^2}{(m_b+m_s)^2}+\frac{3}{2}\right)\,+ \,  \frac{B_4}{B_1}\left(\frac{M_{B_s}^2}{(m_b+m_s)^2}+\frac{1}{6}\right)\left(\eta^{-27/23}-1\right)\right)\biggr]\Bigg|.
\end{split}
\label{BZ}
\end{equation}    
\end{widetext}
In the above eqn. \ref{BZ}, $\eta=\alpha_s(\mu_{NP})/\alpha_s(m_b)$,  SM loop function, $R_{loop}^{SM}=1.310\pm 0.010$ \cite{DiLuzio:2019jyq} and the Bag parameters $B_i,(i=1,4,5)$ are defined in \cite{DiLuzio:2019jyq}. Using eqn. \ref{HBsZ} and \ref{BsBs}, we get:
\begin{equation*}
 \frac{\Delta M_s^{SM+Z'}}{\Delta M_s} \approx |1+ 5\times 10^3\{ (g_L^{bs})^2 +(g_R^{bs})^2 - 9 (g_L^{bs}g_R^{bs}) \} |.
\end{equation*}
The mass of the $Z'$ is significantly higher than the electroweak scale implying that $Z'$ coupling must adhere to $SU(2)_L$ gauge invariance. As a consequence of this $Z'$  interacts with the left-handed neutrinos through the $g_L^{\mu\mu}$ coupling. This interaction introduces an additional term in the effective Hamiltonian which can be represented as:
\begin{equation}
    \mathcal{H}_{eff}^{Z'}\supset  \frac{g_L^{\mu\mu}}{M_{Z'}^2}\,(\bar \nu_{\mu}\gamma^\alpha P_L \nu_{\mu})\,[\bar \mu\gamma_\alpha \,(g_L^{\mu\mu}P_L+ g_R^{\mu\mu}P_R)\,\mu].
\end{equation}
The aforementioned term plays a role in influencing neutrino trident production $(\nu N\xrightarrow{} \nu N\mu^+\mu^-)$ and modifies the cross-section as:\\

\begin{eqnarray}
  R_{\nu} &= &\frac{\sigma}{\sigma_{SM}} \\&  &= \frac{1}{1+(1+4s_W^2)^2} \biggr[\left(1+\frac{\nu^2g_L^{\mu\mu}(g_L^{\mu\mu}-g_R^{\mu\mu})}{M_Z'^2}\right)^2 \nonumber \\ &  &+\left(1+4s_W^2+  \frac{\nu^2g_L^{\mu\mu}(g_L^{\mu\mu}+g_R^{\mu\mu})}{M_{Z'}^2}\right)^2 \biggr]
\label{Rnu}
\end{eqnarray}
Here $\nu =246$ GeV and $s_W= \sin{\theta_W}$. Neutrino trident impose constraints on individual muon coupling $g_{L, (R)}^{\mu\mu}$, experimental measurement for this is $0.82\pm 0.28$. We perform $\chi^2$ analysis to find the NP parameter space allowed by the current flavor data.

\begin{eqnarray}
    \chi^2(g_{L,R}^{bs},g_{L,R}^{\mu\mu}) & = & \sum_K \frac{\mathcal{O}_{K}^{theory}(g_{L,R}^{bs},g_{L,R}^{\mu\mu})-\mathcal{O}_{K}^{exp}}{\sigma^2_{total,K}}.
\end{eqnarray}
where $\mathcal{O}_K^{theory}$ is the theoretical prediction, $\mathcal{O}_{K}^{exp}$ is the experimental central value and  $\sigma_{total,K}$ is the total uncertainty. To calculate $\chi^2$ fit we used constraints from a global fit to all $b\to sl^+l^-$ observables for various NP scenarios as described in Table \ref{WC}.
Contribution of $B_s-\overline{B_s}$ mixing in $\chi^2$ is given as:
\begin{eqnarray}
    \chi^2_{\Delta M_s} &=& \left(\frac{((g_L^{bs})^2+(g_R^{bs})^2-9g_L^{bs}g_R^{bs})-7.69\times {10^{-6}}}{12.94\times{10^{-6}}}\right)^2\nonumber\\
\end{eqnarray}
Where we have used $\Delta M_s^{SM}/\Delta M_s^{exp}=1.04^{+0.04}_{-0.07}$ \cite{DiLuzio:2019jyq}.
The neutrino trident contribution to $\chi^2$ is given by:
\begin{equation}
    \chi^2_{trident} =\left(\frac{R_{\nu}-0.82}{0.28}\right)^2
\end{equation}
The experimental value of $R_{\nu}$ is $0.82\pm 0.28$ \cite{Altmannshofer:2019zhy} and the theoretical expression of $R_{\nu}$ is given by eqn. \ref{Rnu}.\\
The total $\chi^2$ will be:
\begin{equation}
    \chi^2_{total} = \chi^2_{b\to sll}+ \chi^2_{\Delta M_s}+ \chi^2_{trident}
\end{equation}
The permissible new physics couplings are determined by minimizing the $\chi^2$ function. To perform this minimization we used the CERN minimization code MINUIT \cite{James}.
\begin{table*}[tb!]
\centering
\begin{tabular}{ | c| c| c| }
 \hline
  Scenario & Couplings & Wilson Coefficients \\ 

\hline
Z1: $C_{9}^{NP}$ & $g_R^{bs}=0$, $g_L^{bs}$=0.00285,  & $C_9^{NP}= -N_1 g_L^{bs} g_L^{\mu\mu}$\\
& $g_L^{\mu\mu}=g_R^{\mu\mu}=-0.2986$ & \\
\hline
Z2: $C_{9}^{NP}=-C_{10}^{NP}$ & $g_R^{bs}=0$, $g_L^{\mu\mu}= -0.2782$,  & $C_9^{NP}= -(N_1/2) g_L^{bs} g_L^{\mu\mu}$ \\
& $g_R^{\mu\mu}=0$, $g_L^{bs}$=0.00283&\\ 
\hline
Z3: $C_{9}^{NP}=-C_{9}^{'}$ &  $g_L^{\mu\mu}=-0.7125$, $ g_R^{\mu\mu}=-0.712$, & $C_9^{NP}= -(N_1/2) g_L^{bs} (g_L^{\mu\mu}+g_R^{\mu\mu})$ \\
& $g_L^{bs}$= -$g_R^{bs}$ =0.0096 &\\ 
\hline
\end{tabular}
\caption{Universal 1D favored NP scenarios that can be generated in Z' model   }
\label{chi sqr}
\end{table*}
\begin{table*}[htb!]
\centering
\begin{tabular}{ | m{5em} | m{1.5cm}| m{2.5cm}| m{2.5cm} |m{2.5cm}| m{2.5cm}| } 
 \hline
  Angular Observable & Form factor& SM & Z1 &  Z2  & Z3 \\ 
  \hline
  BR$\times 10^7$& LCSR LEET & $1.554\pm0.559$ $1.212\pm 0.605$ & $1.2805\pm0.449$ $1.084\pm0.538$  & $1.231\pm0.444$  $1.022\pm 0.512$ &  $1.149\pm0.394$ $0.991\pm0.488$ \\
  \hline
  $F_{L}$ & LCSR LEET & $0.835\pm 0.016$ $0.721\pm 0.105$ & $0.822\pm0.020$ $0.700\pm0.113$  & $0.8336\pm0.0169$ $0.724\pm0.104$ &$0.812\pm0.023$ $0.667\pm0.121$\\
  \hline
  $A_{FB}$ & LCSR LEET & $-0.015\pm 0.011$  $-0.050\pm 0.026$& $0.0075\pm0.014$ $0.0\pm0.001$ & $-0.0057\pm0.013$ $-0.028\pm0.015$ & $0.00346\pm 0.015$  $-0.008\pm0.004$\\
  \hline
  $P_{1}$ & LCSR LEET& $-0.446\pm 0.149$  $0.374\pm 0.001$ & $-0.3916\pm0.142$ $0.009\pm0.001$ & $-0.412\pm0.146$  $0.225\pm 0.001$& $-0.406\pm0.136$ $0.008\pm0.001$\\
  \hline
  $P_{2}$ & LCSR LEET & $-0.112\pm 0.086$  $-0.171\pm 0.001$& $0.048\pm 0.085$ $0.008\pm0.001$ & $-0.042\pm0.094$ $-0.093\pm0.003$& $0.0205\pm0.088$ $-0.023\pm0.002$\\
  \hline
 $P'_{4}$ &LCSR LEET & $0.975\pm 0.111$  $0.879\pm 0.010$ & $0.935\pm0.110$  $0.840\pm0.011$ & $0.923\pm0.125$ $0.823\pm 0.0165$& $0.932\pm0.104$  $0.832\pm0.012$\\
 \hline
 $P'_{5}$ & LCSR LEET & $-0.349\pm 0.137$  $-0.548\pm 0.0032$ & $-0.149\pm 0.133$ $-0.315\pm 0.001$ & $-0.277\pm 0.143$ $-0.473\pm 0.002$& $-0.206\pm 0.145$ $-0.389\pm 0.0007$\\
 \hline
\end{tabular}
\caption{Angular Observables SM and $Z'$ predictions with 1$\sigma$ for $q^2$ [1-6] GeV$^2$ bin following LCSR and LEET $B\xrightarrow{}K_2^*$ transition form factors \cite{Yang2010qd, Das:2018orb}, for different New Physics scenarios.}
 \label{LEFT+LEET+Z'}
\end{table*}
\begin{figure}[t!]
\centering
\includegraphics[width=0.5\textwidth]{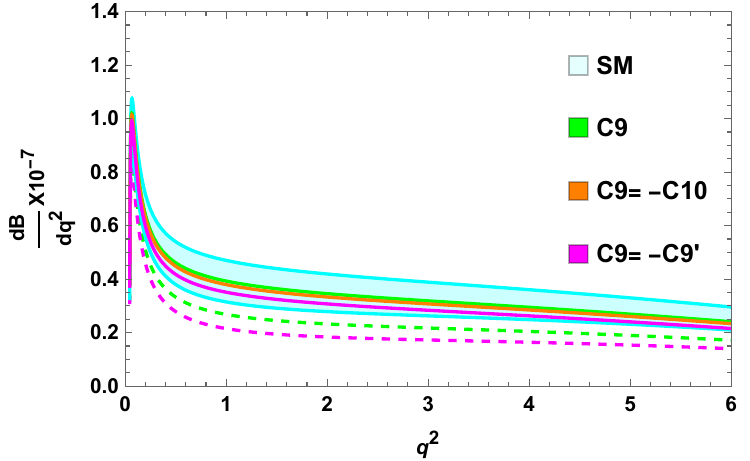}
\caption{Plots for $q^2$ distribution in SM and $Z'$ model along with new physics scenarios for  branching ratio of $B\to K_2^*(\to K\pi)\mu^+\mu^-$ decay.}
\label{fig4}
\end{figure}
 \begin{figure*}[ht! ]
 \includegraphics[width=85mm]{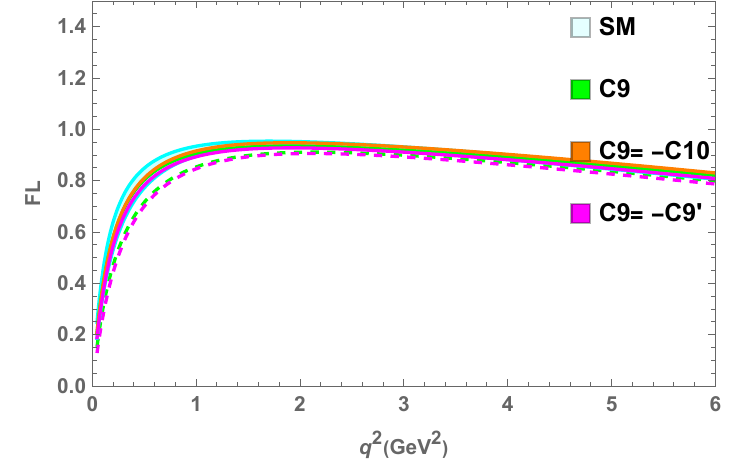}
 \includegraphics[width=85mm]{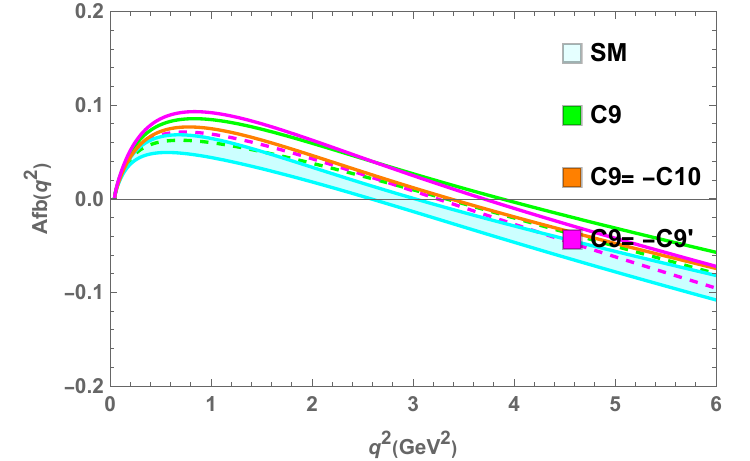}
\caption{Same as for Fig. \ref{fig4} except for $F_L$  and $A_{FB}$observables.}
 \label{fig5}
 \end{figure*}
\begin{figure*}[htb! ]
\includegraphics[width=85mm]{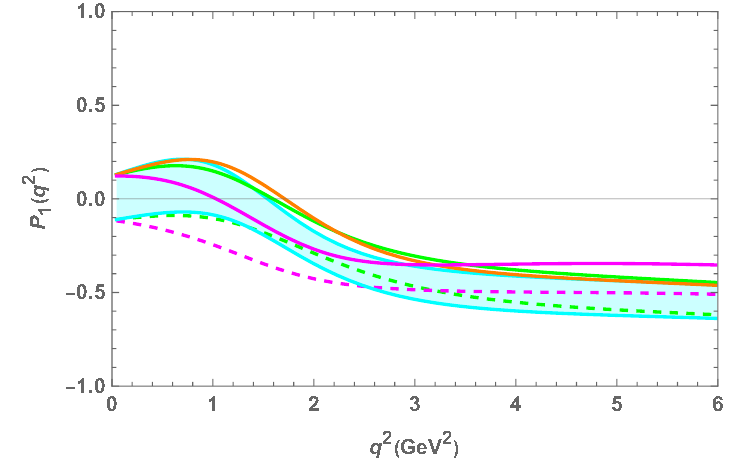}
\includegraphics[width=85mm]{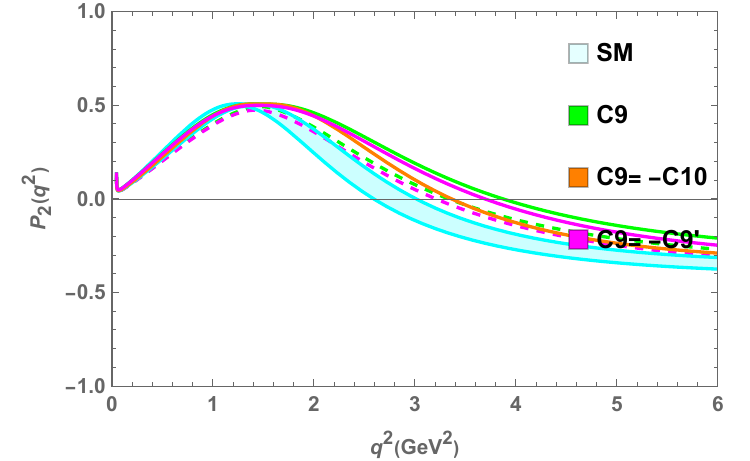}
\\
\includegraphics[width=85mm]{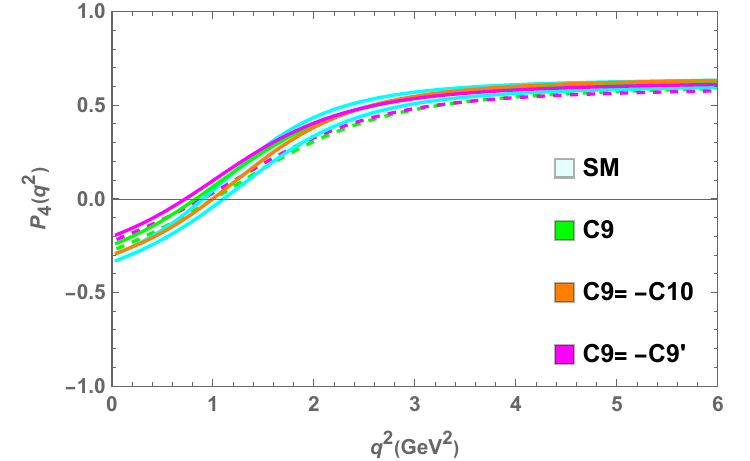}
\includegraphics[width=85mm]{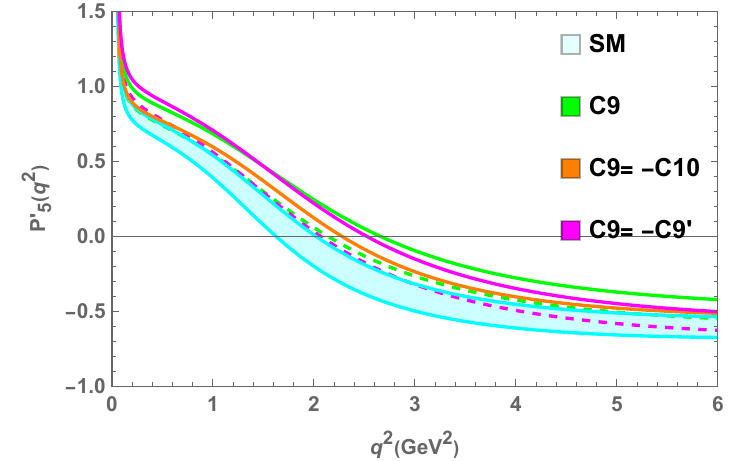}
\\
\caption{ Same as for Fig. \ref{fig4} except for optimized observables$P_i'$.}
\label{fig6}
\end{figure*}
The best-fit values of coupling parameters ($g_{L,R}^{\mu\mu}$, $g_{L,R}^{bs}$) assosicated with the 1TeV  $Z'$ model are given in Table\ref{chi sqr}. Comparing these results with NP Wilson coefficients, Table\ref{WC}, we see that Wilson coefficients within $Z'$ model are consistent with NP Wilson coefficients.
 The $q^2$ dependency for differential branching ratio $\frac{dB}{dq^2}$ in SM along with three NP scenarios in $Z'$ are shown in Fig ~\ref{fig4}. It is lucent from Fig ~\ref{fig4} that there is suppression in $\frac{dB}{dq^2}$ in $Z'$ over the SM. Furthermore, the maximal suppression from SM is about $\sim28\% $ for $Z3$ scenario. From both Longitudinal Polarization plot and Table(\ref{LEFT+LEET+Z'}), it is evident that none of the NP scenarios in $Z'$ shows a significant deviation from SM value in the low $q^2$ bin. Similarly, no remarkable deviation over SM is observed for optimized observable 
$P_4'$ in the $Z'$ model. All three scenarios in $Z'$ model manifest enhancement in optimized observable $P_1$ by $10\sim 20\%$ over SM predictions, and maximum enhancement for Z1 scenario in low $q^2$ bin. In the right pannel of Fig ~\ref{fig5}, the $q^2$ dependency of $A_{FB}$ intimates zero crossing is positive to negative, and zero crossing shifted toward higher $q^2$ than SM value. The prediction of $A_{FB}$ in heavy $Z'$ shows enhancement from SM value. It is apparent from Fig ~\ref{fig4}, that there is a finite enhancement in $P_5'$ in $Z'$ over SM for all scenarios. It leads to maximum enhancement about twofold for Z1 scenario whereas, almost $22\%$ and $42\%$ for Z2 and Z3 scenario for low $q^2$ bin. Enhancement in $P_5'$ in $Z'$ over SM makes it a useful observable in low $q^2$ bin.
\section{conclusion}
In this work, we study several observables for $B\xrightarrow{}K_2^*(\to K\pi)\mu^+\mu^-$ decay in SM, and NP scenarios, under the assumptions that new physics couplings are real and universal in $b\xrightarrow{}s \l^+\l^-$ ($l= e,\mu,\tau$) transition. Firstly, we obtained four-fold angular distribution for $B \xrightarrow{}K_2^*(\to K\pi)l^+l^-$ for vector, axial vector, scalar, and pseudo-scalar interaction, we scrutinize the demeanor of differential branching ratio, forward-backward asymmetry $A_{FB}$, longitudinal polarization asymmetry $F_L$ and optimized observables $P_i^{'}$ in SM and NP scenarios. We explored the above observables by considering all form factors procured through the light cone sum rule and large energy effective theory.\par Our analysis revealed that there is a depletion in the  branching ratio in NP as compared to SM. It is $\sim 15\%$ for S1, and up to $\sim$ 25$\%$ for S2, and S3 NP scenarios. NP solutions provide finite enhancement in muon forward-backward asymmetry $(A_{FB})$. On the other hand, longitudinal fraction $(F_L)$ shows no significant deviations with SM value. We also studied optimized observables $P_i^{(')}$ and observed that there is significant deviations from SM contribution.\par We then consider TeV range non-universal $Z'$ model which can generate 1D favored solutions. The presence of $B_{s}-\Bar{B_s}$ mixing and neutrino trident impose extra constraints on $Z'$ coupling. Using the above constraints, we perform a global fit to determine $Z'$ couplings ($g_{L, R}^{bs}$, $g_{L,R}^{\mu\mu}$), and  Wilson coefficients. We observed that Wilson coefficients under 1TeV $Z'$ model are consistent with new physics Wilson coefficients. We conclude this model does not have any additional advantage over new physics in resolving $b\to s\mu^+\mu^-$ anomalies. 
\section{Appendix}\label{sec: VIII}
\subsection{Effective Wilson Coefficient for $b\xrightarrow{} sl^+l^-$ transition}
The effective Wilson coefficients for $b\xrightarrow{} sll$ transition can be written as describe in \cite{Buras:1994dj}:
\begin{equation*}
  C_7^{eff}= C_7-C_5/3 - C_6
 \end{equation*}
\begin{equation}
\begin{split}
  C_9^{\rm eff}(q^2)& =C_9(\mu)+ h(\hat{m_c},\hat{s})C_0-\frac{1}{2}h(1,\hat{s})(4C_3+4C_4\\ & +3C_5+C_6)- \frac{1}{2}h(0,\hat{s})(C_3+3C_4)\\& +\frac{2}{9}(3C_3+C_4+3C_5+C_6)   
\end{split}
\end{equation}
 With  $\hat{s} = q^2/m_b^2$, $ C_0=C_1+3C_2+3C_3+C_4$ $+3C_5+C_6 $ and $\hat{m_c}$ = $ m_c/m_b$ \\ The auxiliary functions used  as :\\
 if $x=4z^2/s<1$
 \begin{equation*}
 \begin{split}
 h(z,\hat{s})&=-\frac{8}{9}\ln{\frac{m_b}{\mu}} -\frac{8}{9}\ln{z} +\frac{8}{27}+\frac{4x}{9} -\\& \frac{2}{9}(2+x)\lvert {1-x} \rvert ^{1/2}\ln{\lvert \sqrt{1-x}+1/\sqrt{1-x}-1 \rvert}-i\pi  \end{split}
  \end{equation*}
\begin{equation*}
 \begin{split}
 h(z,\hat{s})&= -\frac{8}{9}\ln{\frac{m_b}{\mu}}-\frac{8}{9}\ln{z}+\frac{8}{27}+\frac{4x}{9}\\& -\frac{2(2+x)}{9}\lvert {1-x} \rvert ^{1/2}*2\arctan{\frac{1}{\sqrt{x-1}}},x=\frac{4z^2}{s}>1
 \end{split}
 \end{equation*}
\begin{equation}
     h(0,\hat{s})= -\frac{8}{9}\ln{\frac{m_b}{\mu}}-\frac{4\ln{\hat{s}}}{9} +\frac{8}{27}+\frac{4i\pi}{9}
 \end{equation}
\subsection{Angular Coefficient}

The vector and axial-vector transversity amplitudes can be expressed as:
\begin{equation*}
\begin{split}
A_{0L,R} &= N  \frac{\sqrt{\lambda}}{\sqrt6 m_Bm_{K_2^*}}\frac{1}{2m_{K^*_2}\sqrt {q^2}} [(C_{9-}\mp C_{10-})\\&
((m_B^2-m_{K^*_2}^2-q^2)(m_B+m_{K^*_2})A_1 \\& -\frac{\lambda}{m_B+m_{K^*_2}}A_2) +  2m_b C_7  ( (m_B^2+3m_{K_2^*}^2-q^2)T_2\\& -\frac{\lambda  } {m_B^2-m_{K_2^*}^2}T_3)], \\
A_{\perp L,R} &= -\sqrt{2} N \frac{\sqrt{\lambda}}{\sqrt8m_Bm_{K_2^*}}[(C_{9+}\mp C_{10+})
 \frac{\sqrt \lambda V}{m_B+m_{K^*_2}}\\ &+\frac{2m_b C_{7}}{q^2}\sqrt \lambda T_1], \\
 A_{\parallel L,R} &= \sqrt{2} N\frac{\sqrt{\lambda}}{\sqrt{8} m_B m_{K_2^*}} [(C_{9-}\mp C_{10-}) (m_B+m_{K^*_2}) A_1\\ &+\frac{2m_b C_{7}}{q^2} (m^2_B -m^2_{K^*_2}) T_2],
\end{split}
\end{equation*}
\begin{equation}
A_t= N\frac{\lambda}{\sqrt{q^2}\sqrt{6}m_B m_{K^\ast_2}}\Bigg[2C_{10-}+\frac{q^2}{m_l}\left(C_{P}-C_{P}'\right)\Bigg] A_0,   
\end{equation}

where $C_{9\pm} = (C_{9}  \pm C_{9}^\prime)$, and $C_{10\pm} = (C_{10} \pm C_{10}^\prime)$.
The transversity amplitudes for scalar interactions can be written as

\begin{eqnarray}
  A_S  &= 2 N \sqrt{\lambda}\frac{\sqrt{\lambda}}{\sqrt{6}m_B m_{K^\ast_2}}\Bigg [\frac{(C_{S} - C_{S'})}{(m_b + m_s)} A_0 \Bigg],
\end{eqnarray}
  
where $\lambda = m_B^4+m_{K_2^\ast}^4+q^4-2(m_B^2 m_{K_2^\ast}^2 +m_B^2 q^2 + m_{K_2^\ast}^2q^2)$ 
and normalization constant is given by
\begin{equation*}
 N = \Bigg[ \frac{G_F^2\alpha_e^2}{3\cdot 2^{10}\pi^5m_B^3}|V_{tb}V_{ts}^\ast|^2 \lambda^{1/2}(m_B^2, m_{K_2^\ast}^2, q^2) \mathcal{B}(K_2^\ast\to K\pi)\beta_\ell \Bigg]^\frac{1}{2}\, 
\end{equation*}
\begin{equation}
\beta_\ell = \sqrt{1 - \frac{4m_\ell^2}{q^2}}.    
\end{equation}



\begin{table}[htb!]
\begin{tabular}{|c|c|}
\hline
Input Parameters &  Numeric Value\\
\hline
$m_B$ & 5.279GeV \\
$m_{K_2^*}$ & $1.4324\pm 1.3$ GeV \\
$m_b^{pole}$ & 4.7417GeV\\
$m_c^{pole}$ & 1.5953GeV\\
$|V_{ts}^*V_{tb}|$ &0.04088$\pm$ 0.00055\\
$G_F$ & $1.1663787\times 10^{-5}$\\
$\alpha_e$ & 1/133.28\\
$\tau_{B_s}$ & $1.638\times 10^{-12}$\\
Br$\left(K_2^*\rightarrow K\pi\right)$ & $\left(49.4\pm1.2\right)\%$\\
\hline
\end{tabular}
\caption{The numerical inputs used in our analysis \cite{PDG2018, Detmold:2016pkz} .}
\label{input parameter}
\end{table}



\begin{table}[htb!]
\begin{tabular}{|c|c|c|c|}
\hline
 F & $F^{B_{(s)}K_2^*}(0) $  & $a_{K_2^*}$ & $b_{K_2^*}$\\ 

\hline
V & $0.16\pm 0.02$ & 2.08 & 1.50\\
$A_{0}$ & $0.25\pm 0.04$ & 1.57 & 0.10\\
$A_{1}$ & $0.14\pm 0.02$ & 1.23 & 0.49\\
$A_{2}$ & $0.05\pm 0.02$ & 1.32 & 14.9\\
$T_{1}$ & $0.14\pm 0.02$ & 2.07 & 1.50\\
$T_{2}$ & $ 0.14 \pm 0.02$ & 1.22 & 0.35\\
$T_{3}$ & $0.01^{+0.02}_{-0.01}$ & 9.91 & 276\\
\hline
\end{tabular}
\caption{The relevant form factors parameters ~\cite{Yang2010qd} }
\label{F.F.}
\end{table}
\begin{figure}[t!]
\centering
\includegraphics[width=0.5\textwidth]{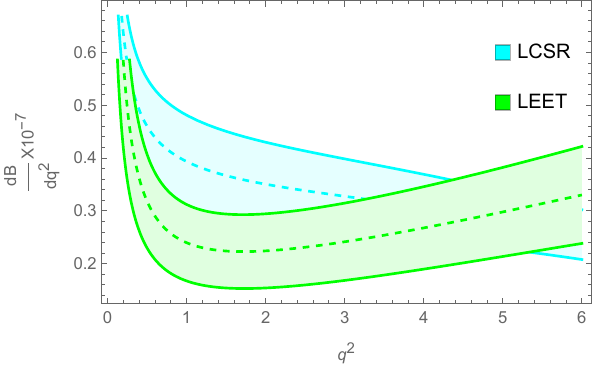}
\caption{Plot for $q^2$ distribution in SM for branching ratio of $B\to K_{2}^{*}(\to K\pi)\mu^+\mu^-$ decay using LCSR and LEET form factor formulation. The solid lines represent the maximum and minimum boundaries for each form factor formulation and dotted lines present cental values.}
\label{fig_SM}
\end{figure}
\section{Acknowledgments}
We thank Dinesh Kunar, Suman Kumbhakar, A. K. Alok and N. R. S. Chundawat for useful discussions. RS acknowledges the financial support from the Science and Engineering Research Board (SERB)
for the National PostDoctoral Fellowship (file no. NPDF/PDF/2021/003328).
\newpage

\end{document}